\documentclass[10pt,onecolumn]{IEEEtran}
\usepackage[numbers,sort&compress]{natbib}
\usepackage{amsmath,amssymb,epsfig}
\usepackage{fancybox}
\usepackage{tikz}
\usetikzlibrary{calc}
\newtheorem{definition}{\bf Definition}
\newtheorem{theorem}{\bf Theorem}
\newtheorem{corollary}{\bf Corollary}

\begin{document}

\bibliographystyle{ieeetr}

\setlength{\parindent}{1pc}

\title{Statistical Properties of Loss Rate Estimators in  Tree Topology}
\author{Weiping~Zhu  }
%\thanks{Weiping Zhu was with University of New South Wales, Australia, email weipingzhu.weiping@gmail.com}}
\date{}
\maketitle

\begin{abstract} Four types of explicit estimators are proposed here to estimate the loss rates of the links in a network with the tree topology and all of them are derived by the maximum likelihood principle. One of the four is developed from an estimator that was used but neglected because it was suspected to have a higher variance. All of the estimators are proved to be either  unbiased or asymptotic unbiased. In addition, a set of formulae are derived to compute the efficiencies and variances of the estimates obtained by the estimators. One of the formulae shows that if a path is divided into two segments, the variance of the estimates obtained for the pass rate of a segment is equal to the variance of the pass rate of the path divided by the square of the pass rate of the other segment. A number of theorems and corollaries are derived from the formulae that can be used to evaluate the performance of an estimator. Using the theorems and corollaries, we find  the estimators from the neglected one are the best estimator for the networks with the tree topology in terms of efficiency and computation complexity.

%that are formed to a partial order. Given the number of estimators and the order, model selection is available for loss tomography since each of the estimators focuses on  an observation.%partial order among them in terms of efficiency or variance.
% where the estimator proposed in \cite{DHPT06} has the largest variance and the least efficiency.
%Further, using the formula an efficient explicit estimator can be easily identified by end-to-end measurement.
\end{abstract}

\begin{IEEEkeywords}
Correlation, Efficiency, Explicit Estimator, Loss Tomography, Maximum Likelihood, Variance.
\end{IEEEkeywords}

\section{Introduction}
\label{section1}
Network characteristics, such as link-level loss rate, delay distribution, available
bandwidth, etc. are valuable information to network
operations, development and researches. Therefore,
 a considerable attention has been given to network
measurement, in particular to  large networks that cross a number of autonomous systems, where security concerns, commercial interests, and administrative
boundary make direct measurement impossible. To overcome the security and administrative obstacles, network tomography was
proposed in \cite{YV96}, where the author suggests the use of
end-to-end measurement and statistical inference to estimate the characteristics of interest.
Since then,
many works have been carried  out to estimate various characteristics that cover loss tomography \cite{CDHT99, CDMT99, CDMT99a, CN00, XGN06, BDPT02,ADV07, DHPT06, ZG05, GW03}, delay tomography
\cite{LY03,TCN03,PDHT02, SH03, LGN06},
 loss pattern tomography \cite{ADV07}, and so on \cite{LHATSBY15,HLATSBY15}. Despite the enthusiasm, there has been a lack of the statistical properties for the loss rate estimators developed for the tree topology, in particular there is no finite sample proporties although some asymptotic properties were presented in \cite{CDHT99, DHPT06}.  Without the properties designated for finite sample, such as efficiency and variance,  it is hard if not impossible to evaluate the performance of an estimator  since all of the estimators proposed so far have the same asymptotic properties. On the other hand, without the properties it is hard to select an estimator for the sample collected from an experiment and it is impossible to determine the number of probes needed for a specific estimating precision.  In order to obtain the properties, we use a different way to model the probing process and then use composite likelihood to reduce the number of correlations  considered in estimation. Finally, we derive a set of  finite sample properties for the estimators proposed in the paper. The finite sample properties are further extended to cover the maximum likelihood estimators (MLE) proposed previously. One of the most important discoveries is a set of formulae to compute the efficiency and variance of the estimates obtained by the estimators proposed and investigated in this paper. 

Using an active method to infer the loss rate of a link, we need to send probing packets, called probes later, from some end-nodes called sources  to another group of nodes called 
receivers and located on the other side of the network, where the paths
connecting the sources to the receivers cover the links of interest. To make statistical inference possible,
 multicast or unicast-based multicast is proposed to send probes from sources to receivers, where an intermediate node is responsible to forward arrived probes to its descendants until the probes reaching their destinations or lost at a node or a link  \cite{HBB00,CN00}.  The lost probes are noticed by the receivers located on the downstream of the link or node that loses the probes. Statistical inference relies on a likelihood function to connect the observations to the loss/pass rates of the links transmitting the probes. An estimating method, such as the maximum likelihood principle, is then applied on the likelihood function to gain a likelihood equation that is also called estimator. The most popular likelihood equation is 
the MLE proposed in \cite{CDHT99}  that is in the form of a polynomial with a degree that is one less than the number of descendants connected to the link of interest \cite{CN00,LHATSBY15}.  If a link has more than 5 descendants, the likelihood equation is a high degree polynomial that requires an iterative procedure, such as  the expectation and maximization (EM) or the Newton-Raphson algorithm, to approximate the solution. However, using an iterative method to estimate loss rates has been widely criticised  for its computational complexity.  Because of this, there has been a persistent effort in the research community to search for an explicit estimator that performs as good as those using iterative approach.  Unfortunately, there has been little progress although a few explicit estimators are proposed.  The few estimators, as others, are evaluated by simulations and the results are far from satisfactory since simulations are neither comprehensive nor conclusive.

  To solve the problems stated above,  we need to use finite sample properties to evaluate the performance of an estimator that requires  a thorough and systematic investigation of the estimators proposed so far. The investigation here is focused on the estimators designated for a network with the tree topology and aims at finding the fundamental principles used by the estimators in estimation and identifying the weaknesses of the estimators. We conduct the investigation and find all of the estimators proposed previously rely on the correlations between predictors and observations to estimate the loss rate of a links. We further finds if a link has $n$ descendants, the MLE proposed in \cite{CDHT99} uses $2^{n}-1$ correlations embedded in the observations of the  descendants inadvertently.  As a result, a high degree polynomial becomes inevitable for the likelihood equation of a link having more descendants. To distinguish the correlations used by the MLE from others, we call them the original correlations. Thus, to have an explicit estimator, we need to reduce the number of correlations used in estimation that can be achieved by either selecting  
  a few correlations from the original ones or creating a few high quality ones. This is because 1) the qualities of the original correlations, measured by the fitness between a predictor and its corresponding observation, are different, some are better than others; and 2) there may have other correlations that are more efficient than the original ones.  This paper is devoted to present the discoveries in the investigation that contribute to loss tomography in four fold.

\begin{enumerate}
%\item The paper unveils such a fact that there could be a number of maximum likelihood estimators (MLE) for the pass rate of a path from the root to an internal node within a network with the tree topology. All of the estimators have identical statistical properties in terms of mean and variance although some of them use a part of the observation obtained in an experiment.
%that an estimator relies on observed correlations to infer the loss rates of a network and the correlations can be decomposed according to observers. Using composite likelihood.we can match a set of the decomposed correlations with its corresponding observation that leads to a large number of explicit estimators.
\item On the basis of composite likelihood \cite{Lindsay88},  three types of estimators: the block wised estimators (BWE), the reduce scaled  estimators (RSE), and the individual based estimators (IBE), are proposed that only use a part of the original correlations.
\item The estimators in BWE and IBE are proved to be unbiased and the estimators in RSE are proved to be  asymptotic unbiased as that proved in   \cite{DHPT06}. A set of formulae are derived for the efficiency and variances of the estimators in RSE and IBE, plus the MLE proposed in \cite{CDHT99}. One of the formulae shows if a path is divided into two segments and we want to estimate the pass rate of a segment from the end to end observation, the variance of the estimates is equal to the variance of the pass rate of the path divided by the square of the pass rate of the path excluding the segment of interest. The formulae also show the weakness of that obtained in \cite{DHPT06}.
\item  The efficiency of the estimators in IBE are compared with each other on the basis of the Fisher information that  shows an estimator using a few observers can be more efficient than another using more and the estimator proposed in \cite{DHPT06} is the least efficient. A similar conclusion is obtained for the estimators in BWE. 
\item The original correlations can be merged into a few that are  much better than the original ones.  Using the merged correlations, we have a set of explicit estimators that perform as good as the MLE proposed in \cite{CDHT99}. 
%\item  Using the formulae, we able to identify an efficient estimator by examining the end-to-end observation that makes model selection not only possible but also feasible. A number of simulations are conducted to verify this feature that also show the connection between efficiency  and robustness of an estimator.
    \end{enumerate}

The rest of the paper is organised as follows. In Section \ref{related work},  we briefly introduce the previous works on the explicit loss rate estimators and point out the weakness of them. In Section \ref{section2}, we introduce the loss model, the notations, and the statistics used in this paper.  Using the model and statistics, we  derive a MLE that considers all of the original correlations  in Section \ref{section3}. We then decompose the original correlations into a number of components and derive a number of likelihood equations for the components in Section \ref{section 4}. In the same section, the original correlations are restructured to a few and a type of estimators based on the few is presented. A statistical analysis of the proposed estimators is presented in Section \ref{section5} that details the statistical properties of the proposed estimators, one of them is the formula to calculate the variances of various estimators. 
%Simulation study is presented in Section \ref{section 6} that compares the performance of five estimators and shows the feasibility of selecting an estimator for a data set.
% A brief discussion of the other features of the proposed estimators, including a number of explicit estimators for a data set with missing data, is also presented in this section.
Section \ref{section7} is devoted to concluding remark.

\section{Related Works}\label{related work}

Multicast Inference of Network Characters (MINC) is the pioneer of using the ideas proposed in \cite{YV96} to estimate the loss rate of a link in a network with the tree topology.  The authors of \cite{CDHT99} use a
Bernoulli distribution to model  the loss behaviours of a link and derive an estimator in the form of a polynomial that is one degree less than the number of descendants
connected to the end node of the path of interest \cite{CDHT99, CDMT99,CDMT99a}.  Apart from that, the authors obtain a number of results  from asymptotic theory, such as the large number behaviour of the estimator and the dependency of the estimator variance on topology. Unfortunately, the results only hold if the sample size $n$ grows indefinitely. In addition, if $n\rightarrow \infty$,  almost all of the estimators proposed previously must have the same results and few can tell the difference between them.  In order to compare the performance of two estimators, experiments and simulation have been widely used but led to little result since there are too many random factors affecting the results obtained from experiments and simulations.

To overcome the problem stated, some simple and explicit estimators, such as that proposed in \cite{ZG05,DHPT06}, are put forward that aims at reducing the complexity of an estimator and hopefully  lead to some insights for further development.
Using this strategy, the authors of \cite{DHPT06} propose an explicit estimator that only considers a correlation from the original ones and claim the asymptotic variance of the estimates obtained by the estimator is the same as that obtained by the estimator proposed in \cite{CDHT99} to first
order. The claim is based on the use of the central limited theorem (CLT) on one of the results acquired by the asymptotic theory in \cite{CDHT99}, where  the covariance between two descendants attached to the path of interest is obtained by assuming the loss rate of a link is very small and then the delta method is used to compute the asymptotic variance on the covariance matrix obtained by the asymptotic theory. The repeated use of the CLT on an estimate makes the claim questionable. Apart from that, some sensitive parameters are cancelled out by approximation. It is easy to prove that under the same condition, most of the estimators proposed so far can achieve at least the same result, if not better,  as that proposed in \cite{DHPT06}.

%The weakness has shown in the example presented in \cite{DHPT06}, where a triple-descendant multicast tree is used  that assumes to have the same pass rate, $\alpha$, for every link. The variance obtained by the proposed method is almost the same as that of the Bernoulli distribution, which, in theory, is only possible if the pass rate of the subtree rooted at the end of the link of interest is 1 or is approaching to 1 that  contradicts to the assumption made for the links.

 % The authors of \cite{DHPT06} use the law of large number to prove the estimate is asymptotic unbiased and use the delta method to prove that if the loss rate of each link is very small, %the variance of the estimates
% is asymptotically the same as that of a MLE, to first order of the loss rate of the path.

In contrast to \cite{DHPT06}, \cite{ADV07} uses an estimator that converts a multicast tree into a binary one and subsequently creates a simple likelihood equation of $A_k$  that is solvable analytically.   Although simulations show the estimator preforms better than that proposed in \cite{DHPT06}, there is little statistical analysis to explain why it is better and the authors even suspect the estimator may yield high variance. Although  the estimator is proved to be a MLE in \cite{Zhu11a}, there is no proof whether it is the same as that proposed in \cite{CDHT99} since the lack of finite sample properties for both of them.

 Although it has been known that finite sample properties are needed to determine the performance of an estimator rather than using simulation and experiment,  there has been little progress that makes the simulation and experiment widely used, even the most recent works presented in \cite{{LHATSBY15,HLATSBY15}}  still rely on simulation to compare the performance of two estimators. 
This paper is devoted to improve the situation and present a few important finite sample properties. 
 
\section{Assumption, Notation and Sufficient Statistics} \label{section2}

 To make the following statistical analysis clear and rigorous, we use  a large number of symbols in the following discussion that may overwhelm the readers who are not familiar with loss tomography. To assist them, the symbols  will be gradually introduced through the paper, where the frequently used symbols will be introduced in the next two sections and the others will be brought up  until needed. In addition, the most frequently used symbols and their meanings are presented in Table \ref{Frequently used symbols and description} for quick reference.

\subsection{Assumption}
\label{assumption}
We assume the probes multicasted from the source to receivers are
independent and network traffic remains statistically stable during the probing process. In addition, the observation obtained at receivers  is considered to be independent identical distributed
($i.i.d.$). Further, the losses occurred at a node or on a link are assumed to be $i.i.d$  as well. 
%almost all of the studies. The study reported in this paper is based on the same assumption.

\subsection{Notation}\label{treenotation}

\begin{figure}
\begin{center}
\begin{tikzpicture}[scale=0.25,every path/.style={>=latex},every node/.style={draw,circle,scale=0.8}]
  \node            (b) at (25,30)  { $v_0$ };
  \node            (d) at (25,24) { $v_1$ };
  \node            (f) at (20,18)  { $v_2$ };
  \node            (g) at (30.5,18) { $v_3$ };
  \node            (j) at (16,12)  { $v_4$ };
  \node            (k) at (23,12) { $v_5$ };
  \node            (l) at (29,12) { $v_6$ };
  \node            (m) at (35,12)  { $v_7$ };
  \node            (r) at (12,6)  { $v_8$ };
  \node            (s) at (18,6) { $v_9$ };
  \node            (t) at (21,6) { $v_{10}$ };
   \node            (u) at (24.5,6)  { $v_{11}$ };
  \node            (v) at (27.5,6)  { $v_{12}$ };
  \node            (w) at (30.5,6) { $v_{13}$ };
  \node            (x) at (33.5,6) { $v_{14}$ };
  \node            (y) at (39,6) { $v_{15}$ };

  \draw[->] (b) edge (d);
  \draw[->] (d) edge (f);
  \draw[->] (d) edge (g);
  \draw[->] (f) edge (j);
  \draw[->] (f) edge (k);
  \draw[->] (g) edge (l);
  \draw[->] (g) edge (m);
  \draw[->] (j) edge (r);
  \draw[->] (j) edge (s);
  \draw[->] (k) edge (t);
  \draw[->] (k) edge (u);
  \draw[->] (l) edge (v);
  \draw[->] (l) edge (w);
  \draw[->] (m) edge (x);
  \draw[->] (m) edge (y);
\end{tikzpicture}
%\centerline{\psfig{figure=graph.eps,width=8.0cm,height=8.0cm}}
\caption{A Multicast Tree} \label{tree example}
\end{center}
\end{figure}
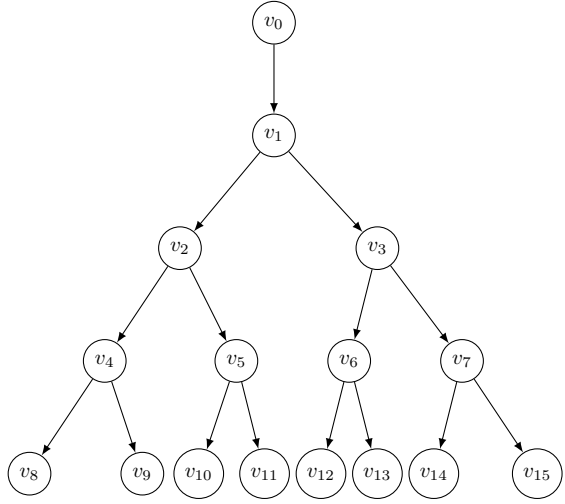

As stated, the network considered in this paper is a multicast tree that is denoted 
by $T=(V, E)$, where  $V=\{v_0, v_1, ... v_m\}$ is a set of
nodes and $E=\{e_1,..., e_m\}$ is a set of directed links that connect the
nodes in $V$.  In addition, $v_k$, and $e_k, k \in \{1,\cdot\cdot, m\}$ are often called node $k$ and link $k$, respectively. By default,  node $0$ is the root node of the multicast tree to which the source is attached. Apart from not having a parent, node $0$ is different from  others by having a single descendant, $v_1$, and using  $e_1$ to connect itself to $v_1$.  In contrast to node $0$, there is a group of nodes called leaf nodes that do not have any descendant. Each leaf node has a receiver attached to it to record the probes received from its parent. Because of this, there is no distinction between a leaf node and the receiver attached to it in the following discussion and $R, R \subset V$ is used to denote them. As a tree , there is one to one correspondence between links and nodes in $T$. If $v_{f(i)}$ is used to denote
the parent of  $v_i$,  $e_i$ is the link connecting $v_{f(i)}$ to
$v_i$.  Figure \ref{tree example} is an example of a multicast binary tree, where nodes are named and connected according to the specification.

A multicast tree, as a tree, can be decomposed into a number of
 multicast subtrees, where $T(k)$ denotes the multicast subtree that has $v_{f(k)}$ as its root and uses $e_k$ to connect $v_f(k)$ to $v_k$. If $v_k \notin R$, $v_k$ connects to a number of multicast subtrees. The nodes directly connected to $v_k$ are called the descendants of node $k$ and denoted by $d_k$. Note that $d_k$ is a nonempty set if $v_k \notin R$. For  the receivers attached to $T(k)$, we use $R(k)$ to denote them. Apart from those, if $x$ is a set, $|x|$ is used to denote the number of elements in $x$. Thus, $|d_k|$ is the number of descendants attached to node $k$ and $|R(k)|$ is the number of receivers attached to $T(k)$. Using the symbols on Figure \ref{tree example}, we have $R=\{v_8, v_9, \cdot\cdot, v_{15}\}$,  $R(2)=\{v_8, v_9, v_{10}, v_{11}\}$,  $d_{2}=\{v_4, v_5\}$,  and $|d_{2}|=2$.

 Although loss tomography aims at estimating the loss rates of the links within a network,  the pass rates of the paths connecting $v_0$ to $v_k, k \in \{1,\cdot\cdot,m\}$ are often estimated instead since there is one to one correspondence between the link-level loss rates and the path-level pass rates. Let $A_k$ denote the pass rate of the path connecting $v_0$ to $v_k$ that is defined as the ratio of the number of probes arrived at node $k$ to the number of probes sent from the source. If we have $A_k, v_k \in V\setminus v_0$, $\alpha_k$,  the pass rate of link $k$, can be obtained by
 \begin{equation}
 \alpha_k=\dfrac{A_k}{A_{f(k)}}. \nonumber
 \end{equation}
Given $\alpha_k$, we are able to compute the loss rate of link $k$ that is equal to $\bar{\alpha}_k=1-\alpha_k$. Because of the correspondence, this paper is focused on estimating $A_k$. 

If $n$ probes are sent from $v_0$ to $R$ in an experiment,
each of them gives rise of an independent realisation
of the passing (loss) process $X$. Let $Z=(x_k^i)^{i=1,...., n}_{v_k\in V}$ donate the states of $T$ in an experiment, where $x_k^i=1$ if probe $i$
reaches $v_k$; otherwise $x_k^i=0$. Among $(x_k^i)^{i=1,...., n}_{v_k\in V}$, only $(x_k^i)^{i=1,...., n}_{v_k\in R}$ are observable and can be used  in estimation. To distinguish the observable states from  others, we call them observations (or sample) later and use
$Y=(y_j^{i})^{i \in \{1,..,n\}}_{v_j \in R}$ to denote them,   where $y_j^i=1$ if probe $i$ is observed by the receiver attached to $v_j$; otherwise, $y_j^i= 0$. Since the probes are multicasted along $T$, we need to isolate the part of the observations related to $A_k$ from $Y$ in estimation.
Let $Y_k=(y_j^{i})^{i \in \{1,..,n\}}_{v_j \in R(k)}$ denote the part of $Y$  that is the complete data set that can be used to estimate $A_k$. $Y_k$ can be further divided into parts if only a part of $Y_k$ is used in estimation, where $Y_k(x)$ denotes the observations obtained by $R(j), v_j \in x \land x \subset d_k$.

\subsection{Problem formulation and statistics}

\label{mlesection}

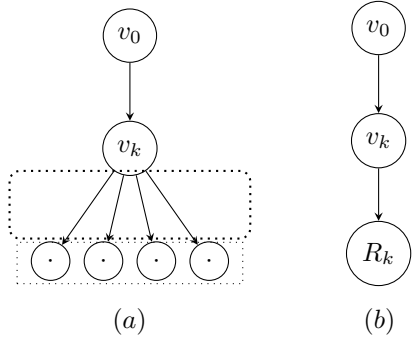
\begin{figure}
\begin{center}
\begin{tikzpicture}[sibling distance=2em,
  every node/.style = {
    align=center,
    top color=white, bottom color=white}]]
  \node[draw,circle] (i) {$v_0$} [grow=south,]
    child { node[draw,circle] {$v_k$} edge from parent[-stealth]
      child { node[draw,circle]  {$\cdot$}}
      child{ node[draw,circle]{$\cdot$}}
      child { node[draw,circle] {$\cdot$}}
      child { node[draw,circle] {$\cdot$} }};
   \draw[dotted]    ($(-1.5,-3.3)$) rectangle ($(1.5,-2.75)$);
    \draw[thick, rounded corners, dotted]    ($(-1.6,-2.7)$) rectangle ($(1.6,-1.8)$);
     \node [below,text width=1cm]     at (0,-3.5){
            $(a)$
        };
\end{tikzpicture}
\begin{tikzpicture}
\draw[] (0, 2.5)  (1, 3.5);
%\draw {latex-latexnew, arrowhead=1cm, line width=1pt}
\end{tikzpicture}
\begin{tikzpicture}
  every node/.style = { 
   align=center,
   top color=white, bottom color=white}]]

 \node[draw,circle] {$v_0$} [grow=south,]
%   \node            (b) at (0,0)  {$0$};
%  \node            (d) at (0, -1) {$K$};
 % \node            (f) at (0,-2)  {$R(k)$};
 % \draw[->] (b) edge (d);
 % \draw[->] (d) edge (f);
  child { node[draw,circle] {$v_k$} edge from parent[-stealth]
    child { node[draw,circle]  {$R_k$}}};
 %   \draw[thick, rounded corners, dotted]    ($(-0.6,-3.4)$) rectangle ($(0.6,-1.8)$);
 \node [below,text width=1cm]     at (0.3,-3.6){
            $(b)$
        };
\end{tikzpicture}
\caption{Transfer a multicast tree to a path of two serially connected links}
\label{abstract tree}
\end{center}
\end{figure}

 In contrast to the previous works that use a special multicast tree isolated from $T$ to formulate a likelihood function of $A_k$,  we change the special multicast tree to a path consisting of two virtual links that are serially connected. The difference between them is illustrated in Figure \ref{abstract tree}. Figure \ref{abstract tree} (a) represents the model used by previous works that uses a virtual link for the path connecting $v_0$ to $v_k$ and a number of virtual links connecting $v_k$ to $R(j), v_j \in d_k$, one for a multicast tree rooted at $v_k$. In contrast,  the model used in this paper is shown in Figure \ref{abstract  tree} (b)  that
 uses a virtual link to replace all of the virtual links connecting $v_k$ to $R(j), v_j \in d_k$, i.e. the links in the dark dot box of Figure \ref{abstract tree} (a), and  uses a node, $R_k$, to replace all of the nodes in the light dot box of Figure \ref{abstract tree} (a). If $\beta_k$ is used to denote the pass rate of the link connecting $v_k$ to $R_k$ and $\gamma_k$ denotes the pass rate from $v_0$ to $R_k$, we have $\gamma_k=A_k\cdot\beta_k,  v_k \in V$. Then, according to the assumption made in Section \ref{assumption},
 the passing process, from $v_0$ to $R(k)$, is a Bernoulli process and a likelihood function about $A_k$ is constructed as follows.
 
  If 
 \begin{equation}
 yk^i=\bigvee_{\substack{v_j \in R(k)}} y_j^i \nonumber
 \end{equation} 
is defined as the observation of $R(k)$ for probe $i$, 
 \begin{equation}
n_k(d_k)=\sum_{i=1}^n yk^i
\label{nk1}
\end{equation}
becomes a statistic of $Y_k$. If $\hat\gamma_k$ is used for the empirical value of $\gamma_k$,  we have $\hat\gamma_k=\dfrac{n_k(d_k)}{n}$. Note that $\hat\gamma_j=\dfrac{n_j(j)}{n}, v_j \in R$ is the empirical pass rate of the path from the root to node $j$.
Given all of those, i.e. the assumptions made in  Section \ref{assumption}, the definitions presented in Section \ref{treenotation}, and the model specified above, the joint distribution of the passing process for $Y_k$ can be written as 

\begin{equation}
{\cal P}(Y_k=\{yk^1,yk^2,\cdot\cdot, yk^n\}, n_k(d_k))= (A_k\beta_k)^{n_k(d_k)}(1-A_k\beta_k)^{n-n_k(d_k)}.
\label{prob function}
\end{equation}
Accordingly, a likelihood function of $A_k$ for observation $Y_k$ and statistic $n_k(d_k)$ is written as follows:
\begin{equation}
{\cal L}(A_k, Y_k, n_k(d_k))=(A_k\beta_k)^{n_k(d_k)}(1-A_k\beta_k)^{n-n_k(d_k)}.
\label{likelihood function}
\end{equation}
We can then prove  $n_k(d_k)$ is a sufficient statistic with respect to ({\it wrt.}) the passing process of $A_k$ for the observation obtained by $R(k)$. Rather than using the well known factorisation theorem in the proof, we directly use the mathematic definition of a sufficient statistic (See definition 7.18 in \cite{RM96}) to achieve this and present it as a theorem.

\begin{theorem}\label{complete minimal sufficient statistics}
Let $Y_k$ be the i.i.d random sample obtained by $R(k)$ from the probes sent by the source and governed by a Bernoulli process as (\ref{prob function}). The statistic $n_k(d_k)$ is minimal
sufficient in respect of the observation of $Y_k$.
\end{theorem}

\begin{IEEEproof}
According to the definition of sufficiency, we need to prove
\begin{equation}
{\cal P}(Y_k=\{yk^1,yk^2,\cdot\cdot, yk^n\}|n_k(d_k)=t)=\dfrac{{\cal P}(Y_k=\{yk^1,yk^2,\cdot\cdot, yk^n\}, n_k(d_k)=t)}{{\cal P}(n_k(d_k)=t)} \nonumber
\end{equation} is independent of $A_k$.

Given (\ref{prob function}), the passing process with observation of $n_k(d_k)=t$ is a random process that yields the binomial distribution as follows
\[
{\cal P}(n_k(d_k)=t)=\binom{n}{t}(A_k\beta_k)^{t}(1-A_k\beta_k)^{n-t}.
\]
Then, we have
\begin{eqnarray}
{\cal P}(Y_k=\{yk^1,yk^2,\cdot\cdot, yk^n\}|n_k(d_k)=t)&=&\dfrac{(A_k\beta_k)^{t}(1-A_k\beta_k)^{n-t}}{\binom{n}{t}(A_k\beta_k)^{t}(1-A_k\beta_k)^{n-t}.
} \nonumber \\
=\dfrac{1}{\binom{n}{t}}, \nonumber
\end{eqnarray}
which is independent of $A_k$ and $\beta_k$. Then, $n_k(d_k)$ is a sufficient statistic.

Apart from the sufficiency,
 $n_k(d_k)$,  as defined in (\ref{nk1}), is a count of the probes reaching $R(k)$ that counts each probe once and once only regardless of how many receivers observe the probe. Therefore,  $n_k(d_k)$ is a minimal sufficient statistic in regard to the observation of $R(k)$.
\end{IEEEproof}

\subsection{Statistics considering a part of observation}
\label{mlestatistics}
Apart from using $n_k(d_k)$, there are other statistics that can be used  to estimate $A_k$. Some of them only use a part of the observation in $Y_k$ that counts the number of probes reaching a particular group of receivers. Let $ x, x \subset d_k \land |x|\geq 2$ be a subset of the multicast subtrees rooted at node $k$ that have $\{R(j): v_j \in x\}$ attached. Then, we have a path as Figure \ref{abstract tree}  (b) and a likelihood function as (\ref{likelihood function}), where $n_k(d_k)$ is replaced by: 
 \begin{equation}
 n_k(x)=\sum_{i=1}^n yz^i.
 \label{nk2}
 \end{equation}
 where
 \begin{equation}
 yz^i=\bigvee_{\substack{v_j \in R(z)\\ v_z \in x}} y_j^i. \nonumber
\end{equation} 
\noindent  If $N_k$ is used to denote the number of probes reaching node $k$, we have $N_k \geq n_k(d_k)\geq n_k(x), x \subset d_k$. Accordingly, $\beta_k(x), x \subset d_k$ is used to denote the pass rate of the multicast subtrees consisting of $T(j), v_j \in x$. Given $n_k(x)$ and $\beta_k(x)$, we can  write a likelihood function of $A_k$ and use the same procedure as that in Section \ref{mlesection} to prove $n_k(x)$ a sufficient statistic in the context of the observation obtained by $\{R(j): v_j \in x\}$. In addition, an estimator based on $n_k(x)$ can be created. If $n_k(x) =n_k(d_k)$ or $n_k(x) \approx n_k(d_k)$, the estimator derived from $n_k(x)$ is expect to perform as good as that uses $n_k(d_k)$ and $\beta_k$.

\begin{table}[htp]
\caption{Frequently used symbols and description}
\begin{center}
\begin{tabular}{|c|l|} \hline
Symbol & Desciption \\\hline
$T(k)$ & the subtree rooted at link $k$. \\ \hline
$d_k $& the descendants attached to node $k$. \\\hline
$R(k)$ & the receivers attached to $T(k)$. \\ \hline
$A_k$ & the pass rate of the path from $v_0$ to $v_k$. \\ \hline
$\beta_k$& the pass rate of the subtree rooted at node $k$. \\ \hline
$\beta_k(x)$& the pass rate of the subtree consisting of $T(j), v_j \in x \land x \subset d_k$. \\\hline
$\gamma_k$& $A_k*\beta_k$, pass rate from $v_0$ to $R(k)$, via $v_k$. \\ \hline
$N_k$ & the number of probes reaching node $k$. \\ \hline
%$C_k$ & the subtrees attached to node $k$ \\ \hline
%$C_k(x)$ & the subset of $C_k$, where $x$ is a subset of $d_k$ \\ \hline
$x_k^i$ & the state of $v_k$ for probe $i$.  \\ \hline
$\sum_k$ & the $\sigma$-algebra created from $d_k$. \\ \hline
$n$ & the number of probes sent in an experiment, \\ \hline
$n_k(d_k)$ & the number of probes reaches $R(k)$. \\ \hline
$n_k(x)$ & the number of probes reaches the receivers attached to $T(j), v_j \in x$. \\ \hline
$I_k(x)$ & the number of probes observed by the members of $x$. \\ \hline
$Y$ &  the observation obtained in an experiment. \\ \hline
$Y_k, v_k \in V$& the part of $Y$ obtained by $R(k)$. \\ \hline
$Y_k(x), x \subset d_k$&  the part of $Y$ obtained by $R(j), v_j \in x$.\\ \hline
\end{tabular}
\end{center}
\label{Frequently used symbols and description}
\end{table}

\section{Estimator Analysis} \label{section3}
This section is dedicated to the analysis of the MLE that considers all of the original correlations. By the analysis, we are able to identify all of the predictors and the corresponding observations used in the MLE and find the connections between them.

\subsection{Maximum Likelihood Estimator based on Original Correlations} \label{2.a}

Turning the likelihood function presented in (\ref{likelihood function}) into a log-likelihood function, we have
 \begin{equation}
 \log {\cal L}(A_k, Y_k, n_k(d_k))=n_k(d_k)\log (A_k\beta_k)+(n-n_k(d_k))\log(1-A_k\beta_k).
 \label{likelihood}
 \end{equation}
 Differentiating (\ref{likelihood}) {\it wrt.} $A_k$
and letting the derivatives be 0, we have

\begin{eqnarray}
\dfrac{n_k(d_k)}{A_k}-\dfrac{(n-n_k(d_k))\beta_k}{1-A_k\beta_k}=0,
\label{likelihood equation}
\end{eqnarray}
and then
\begin{eqnarray}
A_k\beta_k&=&\frac{n_k(d_k)}{n}.
\label{AkBk}
\end{eqnarray}
Since neither $A_k$ nor $\beta_k$ can be solved  directly from (\ref{AkBk}), we need to use a connection between $A_k$ and $\beta_k$  to derive the MLE.
Given the {\it i.i.d.} model assumed previously and the multicast used in probing,  the following equation is used to link $A_k$ to $\beta_k$
\begin{equation}
1-\beta_k=\prod_{v_j \in d_k} (1-\dfrac{\gamma_j}{A_k}).
\label{beta-k}
\end{equation}
Solving $\beta_k$ from (\ref{beta-k}) and using it in (\ref{likelihood equation}), we have a MLE as
\begin{equation}
1-\dfrac{n_k(d_k)}{n \cdot A_k}=\prod_{v_j \in d_k} (1-\dfrac{\gamma_j}{A_k}).
\label{realmle1}
\end{equation}
Using $\gamma_k$ to replace $\dfrac{n_k(d_k)}{n}$ since the latter is the empirical value of the former, we have a likelihood equation as follows:
\begin{equation}
1-\dfrac{\gamma_k}{A_k}=\prod_{v_j \in d_k} (1-\dfrac{\gamma_j}{A_k})
\label{minc}
\end{equation}
that is identical to that proposed in \cite{CDHT99}.
 %In order to find the correlations considered in the MLE, we use (\ref{realmle1}) rather than (\ref{minc}) in the rest of this section because it explicitly connects observations to correlations, i.e. $n_k(d_k)$ to $\prod_{v_j \in x} \gamma_j, x \in d_k$.

\subsection{Predictor and  Observation}
%
%(\ref{beta-k}) not only plays a key role in the derivation of the likelihood equation, but also an important role in the analysis.  The equation shows the connection between the parameter to be estimated and the correlations to be considered. If all of the correlations among the descendants, from simple to complex,  are considered, we have a likelihood equation as
%\begin{equation}
%1-\dfrac{n_k(d_k)}{n \cdot A_k}=\prod_{v_j \in d_k} (1-\dfrac{\gamma_j(d_j)}{A_k}),
%\label{realmle}
%\end{equation} that is called a full likelihood equation in literature. Selecting $D, D \subseteq d_k \mbox{ and } \#(D) \geq 2$ in (\ref{realmle1}), we can have a large number of likelihood equations and some of them can be solved analytically.  Given this, what we need to do is to find all of the correlations that can be used in estimation, isolate the corresponding observation for each correlation, create estimators and evaluate them.
%
To make the correlations involved in (\ref{realmle1}) visible,  the left hand side (LHS) and the right hand side (RHS) of  (\ref{realmle1}) are expanded, where the terms obtained from the LHS are observations and the terms from the RHS  are predictors. There is one to one  correspondence between the terms on the two sides. Each of them is called a correlation and there are $2^{|d_k|}-1$ correlations that are called the original correlations.  For instance, $\gamma_i\cdot \gamma_j/A_k, v_i, v_j \in d_k \land i \neq j$, is the predictor of the probes simultaneously observed by the receivers attached to subtree $i$ and subtree $j$, i.e.  the number of probes observed by at least a receiver from each of the subtrees.

To represent the original correlations,
 a $\sigma$-algebra, $S_k$, is created over $d_k$ and  let $\Sigma_k=S_k \setminus \emptyset$ be the non-empty sets in $S_k$. Each member in $\Sigma_k$ corresponds to a pair of a predictor and its observation.  If  the number of elements in a member of $\Sigma_k$ is defined as the degree of the correlation, $\Sigma_k$ can be divided into $|d_k|$ exclusive groups, one for
 a degree of the correlations that vary from 1 to $|d_k|$. Let $S_k(i), i \in \{1,\cdot\cdot,|d_k|\}$ denote the group of  correlations that are all $i$ degree. For example, if $d_k=\{i,j,k,l\}$, $S_k(2)=\{(i,j),(i,k),(i,l),(j,k),(j,l),(k,l)\}$ denotes the pairwise correlations in $d_k$, and $S_k(3)=\{(i,j,k),(i,j,l),(i,k,l),(j,k,l)\}$ denotes the triplet-wise correlations.

Given $\Sigma_k$,  $n_k(d_k)$ can be decomposed into the probes that are observed simultaneously by the members of  $\Sigma_k$. The simultaneous observation by the member of   $x, x \in \Sigma_k$ and $|x|>1$, is defined as if $\forall j, v_j \in x$ there is at least a receiver attached to $T(j)$ observes the probe.
 To explicitly express $n_k(d_k)$ by $n_j(d_j), v_j \in d_k$,  $I_k(x), x \in \Sigma_k$ is introduced to return  the number of probes observed simultaneously by  the members of $x$ in an experiment.
Let $u_j^i$ be the observation of $R(j)$ for probe $i$ that is defined as:
\[
u_j^i=\bigvee_{v_k \in R(j)} y_k^i,
\]
then
\begin{equation}
I_k(x)=\sum_{i=1}^n \bigwedge_{v_j \in x} u_j^i, \mbox{\vspace{1cm} } x \in \Sigma_k.
\label{I-k x}
\end{equation}
If $x=(j)$,
\[
I_k(x)=n_j(d_j), v_j \in d_k,
\]
Given the above, $n_k(d_k)$ can be written as:

\begin{equation}
n_k(d_k)=\sum_{i=1}^{|d_k|}(-1)^{i-1}\sum_{x \in S_k(i)} I_k(x) \label{n_k value}
\end{equation}
according to  the inclusion-exclusion principle \cite{AS10} that ensures each probe observed by $R(k)$ is counted once and once only in $n_k(d_k)$.

\subsection{Correlations between Predictors and Observations}

Given  (\ref{n_k value}), we are able to prove the MLE proposed in \cite{CDHT99} aims at minimising the difference between the predictors and the observations among  the original correlations and have the following theorem.

\begin{theorem} \label{minctheorem}

\begin{enumerate}
\item (\ref{realmle1}) is a full likelihood estimator that considers all of the correlations in $\Sigma_k$;
\item (\ref{realmle1}) consists of  observations and predictors, one for a member of $\Sigma_k$; and
\item the estimate obtained from (\ref{realmle1})  is a fit that aims at minimising the alternating differences between observations and predictors.
    \end{enumerate}
\end{theorem}
\begin{IEEEproof}
(\ref{realmle1}) is a full likelihood estimator that considers all of the correlations in $\Sigma_k$. To prove 2) and 3),  we expand the both sides of (\ref{realmle1}) and pair the observations with the  predictors  in $S_k$. There are three steps to achieve them.
\begin{enumerate}
\item  If  we use (\ref{n_k value}) to replace $n_k(d_k)$ from LHS of (\ref{realmle1}),  the LHS becomes:
\begin{equation}
1-\dfrac{n_k(d_k)}{n\cdot A_k}= 1-\dfrac{1}{n\cdot A_k}\big
[\sum_{i=1}^{|d_k|}(-1)^{i-1}\sum_{x \in S_k(i)}I_k(x)].
\label{nkexpansion}
\end{equation}
\item If we expand the product term located on  the RHS of (\ref{realmle1}), we have:
\begin{equation}
\prod_{v_j \in d_k}(1-\dfrac{\gamma_j}{A_k})=1-\sum_{i=1}^{|d_k|}(-1)^{i-1}\sum_{x \in S_k(i)}\dfrac{\prod_{v_j \in x} \gamma_j}{A_k^i} \label{prodexpansion}
\end{equation} where the alternative adding and subtracting operations intend to remove the impact of redundant observation.
\item Deducting 1 from both (\ref{nkexpansion}) and (\ref{prodexpansion})
and then multiplying the results by $A_k$, (\ref{realmle1}) turns to
\begin{eqnarray}
\sum_{i=1}^{|d_k|}(-1)^{i}\sum_{x \in S_k(i)}\dfrac{I_k(x)}{n}
=\sum_{i=1}^{|d_k|}(-1)^{i}\sum_{x \in S_k(i)}\dfrac{\prod_{v_j \in x} \gamma_j}{A_k^{i-1}}. \label{statequal}
\end{eqnarray}
It is clear there is one to one correspondence between the terms across
the equal sign, where the terms on the LHS are the observations and the terms on the RHS are the predictors. If we rewrite
(\ref{statequal}) as
\begin{equation}
\sum_{i=1}^{|d_k|}(-1)^{i}\sum_{x \in S_k(i)}\Big(\dfrac{I_k(x)}{n} -\dfrac{\prod_{v_j \in x}
\gamma_j}{A_k^{i-1}}\Big)=0, \label{correspondence}
\end{equation}
\end{enumerate}
the correspondence becomes obvious.
\end{IEEEproof}
(\ref{correspondence}) shows that the MLE is a polynomial of $A_k$ and the degree of the polynomial is determined by $|d_k|$. To distinguish the MLE from others, we call it original MLE in the rest of the paper.

%Theorm \ref{minctheorem}
%shows that the counting process that ensures each probe is only counted once regardless of how many receivers observe it corresponds to such a process that eliminates various correlations in modeling.

\section{Explicit Estimators based on Composite Likelihood} \label{section 4}
(\ref{correspondence}) shows that the original MLE takes into account all of the correlations in $\Sigma_k$. If the number of subtrees rooted at node $k$ is larger than 5, the estimator is a high degree polynomial that could not be solved analytically  according to Galois theory. To have an explicit estimator in such a circumstance, we need to reduce the number of correlations considered in estimation and there are a number of strategies to achieve this. We here propose three of them that use composite likelihood, which is also called
pseudo-likelihood by Besag in \cite{Besay74}, to structure likelihood functions. The three are named  reduce scaled, block-wised, and individual based, respectively. The reduce scaled strategy, as named, is a down-size version of the original MLE that removes a  number of subtrees rooted at node $k$ from consideration and then uses the maximum likelihood principle on the rest to estimate $A_k$. The block-wised strategy differs from the reduce scaled one by dividing the original correlations into a number of blocks, one for a degree of correlations, from pairwise to $d_k$-wise. The individual based one,  in contrast to the other two, considers a correlation at a time that leads to a large number of estimators. 

Apart from the three, another is developed from the alternative estimator used in \cite{ADV07} that has been neglected because the authors suspect the estimates obtained by the estimator have a high variance. This type of estimators differs from the above by merging all of the correlations into a few more efficient ones that challenges the claim made at the beginning of last paragraph and shows the degree of an estimator should be independent to the number of the descendants attached to the path of interest.

\subsection{Reduce Scaled Estimator (RSE)}

Rather than considering all of the correlations in $\Sigma_k$, the correlations can be divided into groups according to the subtrees rooted at node $k$. Let $x, x \subset d_k$ be the group to be considered by an estimator in RSE. The log-likelihood function considering the correlations among $x$ is as follows:
 \begin{equation}
\log {\cal L}(A_k, Y_k(x), n_k(x))= n_k(x)\log (A_k\beta_k(x))+(n-n_k(x))\log(1-A_k\beta_k(x))
 \label{likelihood RSE}
 \end{equation}
 where $n_k(x)$ as defined in Section \ref{mlestatistics} is the number of probes reaching node $k$ confirmed from the observations of the receivers attached to $\{T(j): v_j \in x\}$. If
 $\beta_k(x)$ denotes the pass rate of  the virtual link consisting of $\{T(j): v_j \in x\}$, the following connects $\beta_k(x)$ to $A_k$ 
\begin{equation}
1-\beta_k(x)=\prod_{v_j \in x} (1-\dfrac{\gamma_j}{A_k}). \nonumber
\end{equation}
Then, a similar likelihood equation as (\ref{realmle1}) is obtained and presented as follows:
\begin{equation}
1-\dfrac{n_k(x)}{n \cdot A_k}=\prod_{v_j \in x} (1-\dfrac{\gamma_j}{A_k}). \nonumber
\end{equation}
If $|x|\leq 5$, the equation is solvable analytically. The estimators in RSE are denoted by $Am_k(x), x \subset d_k$.

\subsection{Block-wised Estimator (BWE)}

(\ref{correspondence}) shows that the correlations involved in the original MLE can be divided into $|d_k|-1$ blocks, from pairwise to $|d_k|$-wise.  Each of them can be written as a likelihood function. In order to use a unique likelihood function for all of them, we let the likelihood function considering single correlation be 1. Then, the likelihood function considering $i$-wise correlations is denoted as $L_c(i; A_k; y)$ that can be expressed uniformly.

\begin{definition} \label{recursive corollary}
There are a number of composite likelihood functions, one for a degree of correlations, varying from pairwise to $|d_k|$-wise. The composite likelihood function  $L_c(i; A_k; y), i \in \{2,\cdot\cdot, |d_k|\}$ has a  form as follows:

\begin{eqnarray}
L_c(i; A_k; y) &=&\dfrac{\prod_{x \in S(i)} (A_k\beta_k(x))^{n_k(x)}(1-A_k\beta_k(x))^{n-n_k(x)}}{\prod_{x' \in S(i-1)}(A_k\beta_k(x'))^{n_k(x')}(1-A_k\beta_k(x'))^{n-n_k(x')}},  i \in \{2,\cdot\cdot,|d_k|\}.
\label{recursive form}
%L_c(i; A_k; y) &=&\prod_{x \in S_k(i)}(A_k\psi_k(x))^{n_k(x)}(1-A_k\prod_{x \in S_k(i)}\psi_k(x))^{n-n_k(x)} \label{recursive form} \\
%&& i \in \{2,\cdot\cdot,|d_k|\} \nonumber
\end{eqnarray}
 \end{definition}
 Let $A_k(i)$ be the estimator derived from $L_c(i; A_k; y)$. Then,  we have the following theorem.
\begin{theorem} \label{all explicit}
Each of the composite likelihood
equations obtained from (\ref{recursive form}) is an explicit estimator of $A_k$ that is as follows:
\begin{equation}
A_k(i)=\Big (\dfrac{\sum_{\substack{ x \in S_k(i)}}
\prod_{v_j \in x} \gamma_j}{\sum_{x \in S_k(i)}\dfrac{I_k(x)}{n}}{\Big )} ^{\frac{1}{i-1}},  i \in
\{2,.., |d_k|\}. \label{approximateestimator}
\end{equation}
\end{theorem}
\begin{IEEEproof}
Firstly, we can write (\ref{recursive form}) into a log-likelihood function and then differentiate the function {\it wrt} $A_k$. As (\ref{AkBk}), we cannot solve $A_k$ or $\beta_k(x)$ directly from the derivative and we need to consider other correlations as (\ref{beta-k}). We then have an equation as
\begin{equation}
\frac{\partial\log L_c(i,A_k; y)}{\partial A_k}=\sum_{x \in
S(i)} \Big [1-\dfrac{\gamma_k(x)}{A_k}-\prod_{q\in x}(1-\dfrac{\gamma_q}{A_k})\Big ]-\sum_{x' \in S(i-1)} \Big [1-\dfrac{\gamma_k(x')}{A_k}-\prod_{q\in x'}(1-\dfrac{\gamma_q}{A_k})\Big ]. \nonumber
\end{equation} The two summations can be expanded as (\ref{realmle1}) and only the terms related to  i-wise correlations left  since all other terms in the first summation are canceled by the terms of the second summation. The likelihood equation as (\ref{approximateestimator}) follows.
\end{IEEEproof}
In the rest of the paper, $A_k(i)$ is used to refer to the $i-wise$ estimator and $\widehat A_k(i)$ refers to the estimate obtained by $A_k(i)$.

\subsection{Individual based Estimator (IBE)}

Instead of considering a block of correlations together, we can consider a correlation at a time that results in a large number of estimators. Each of them has a similar likelihood function as (\ref{likelihood RSE}), where $\beta_k(x)$ and $n_k(x)$ are replaced by $\psi_k(x)$ and $I_k(x)$, respectively. $\psi_k(x)=\prod_{v_j \in x} \alpha_j\beta_j, x \subseteq d_k$, is the pass rate of $\{T(j): v_j \in x\}$. If $\Sigma_k'=\Sigma_k \setminus S_k(1)$ is the correlations considered by IBE, the log-likelihood function for $A_k$ given observation $I_k(x)$ is equal to
 \begin{eqnarray}
{ \cal L}(A_k, Y_k(x), I_k(x))= I_k(x)\log (A_k\psi_k(x))+(n-I_k(x))\log(1-A_k\psi_k(x)),  \mbox{   } x \in \Sigma_k'.
 \label{Al likelihood1}
 \end{eqnarray}
 We then have the following theorem.
\begin{theorem}  \label{local estimator}
Given (\ref{Al likelihood1}), $A_k\psi_k(x)$ is a Bernoulli process. The MLE for $A_k$ given $I_k(x)$ equals to
\begin{equation}
Al_k(x)=\Big(\dfrac{\prod_{v_j\in x} \gamma_j}{\dfrac{I_k(x)}{n}}\Big)^{\frac{1}{|x|-1}}.  \mbox{   }    x \in \Sigma_k'
\label{local estimator1}
\end{equation}
\end{theorem}
	\begin{IEEEproof} Using the same procedure as that used in Section \ref{2.a}, we have the theorem.
\end{IEEEproof}

Comparing (\ref{approximateestimator}) with (\ref{local estimator1}), we can find that $\widehat Al_k(x)$, where $|x|=i$, is a type of geometric mean and $\widehat A_k(i)$ is the arithmetic mean of $\widehat Al_k(x), x \in S_k(i)$. Therefore, $A_k(i)$ is more robust than $Al_k(x)$.

\subsection{Estimator based on Restructure}

Since the publication of \cite{CDHT99}, few has questioned such a claim made at the beginning of this section, i.e. the computation complexity of the original MLE is related to the number of descendants connected to the link of interest. Unfortunately, this claim is incorrect that is due to the use of (\ref{beta-k})  in the derivation of the original MLE. If (\ref{beta-k}) is replaced by a low degree polynomial to connect $A_k$ to $\beta_k$, we can have a group of explicit MLEs that perform as good as the original MLE.   This group of estimators are called the merged MLE since the observations of the receivers need to be merged in a different way than that used in \cite{CDHT99}.

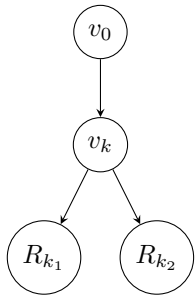
\begin{figure}
\begin{center}
\begin{tikzpicture}
%\draw[] (0, 2.5)  (1, 3.5);
%\draw {latex-latexnew, arrowhead=1cm, line width=1pt}
\end{tikzpicture}
\begin{tikzpicture}
  every node/.style = { 
   align=center,
   top color=white, bottom color=white}]]

 \node[draw,circle] {$v_0$} [grow=south,]
%   \node            (b) at (0,0)  {$0$};
%  \node            (d) at (0, -1) {$K$};
 % \node            (f) at (0,-2)  {$R_k$};
 % \draw[->] (b) edge (d);
 % \draw[->] (d) edge (f);
  child { node[draw,circle] {$v_k$} edge from parent[-stealth]
    child { node[draw,circle]  {$R_{k_1}$}}
    child { node[draw,circle]  {$R_{k_2}$}}};
 %   \draw[thick, rounded corners, dotted]    ($(-0.6,-3.4)$) rectangle ($(0.6,-1.8)$);
\end{tikzpicture}
\caption{Transfer a multicast tree to a binary tree}
\label{binary tree}
\end{center}
\end{figure}

To restructure the multicast subtrees rooted at node $k$,   the merged MLE divides the  multicast subtrees rooted at node $k$ into exclusive groups and considers a group as a virtual link. For example in Figure \ref{binary tree}, all of the multicast trees rooted at node $k$ can be divided into two groups: $k_1$ and $k_2$, to maximally reduce the number and degree of the correlations to be considered in estimation. It then uses (\ref{nk2}) to
merge the observation of the subtrees in $k_1$ and $k_2$, respectively, and use them as the probes reaching $R_{k1}$ and $R_{k2}$.
Let $n_k(k_1)$ and $n_k(k_2)$ be the numbers of probes observed by $R_{k1}$ and $R_{k2}$, respectively. Then, if we use 
\[
1-\beta_k=\prod_{j \in \{k_1,k_2\}}(1-\dfrac{\gamma_j}{A_k}),
\]
to replace (\ref{beta-k}) in the derivation of the original MLE, we have a MLE as follows:
\begin{equation}
1-\dfrac{\hat \gamma_k}{A_k}=(1-\dfrac{\hat \gamma_{k_1}}{A_k})(1-\dfrac{\hat \gamma_{k_2}}{A_k})
\label{linear equation}
\end{equation}
where $\hat\gamma_k=\dfrac{n_k(d_k)}{n}$,
$\hat\gamma_{k_1}=\dfrac{n_k(k_1)}{n}$, and
$\hat\gamma_{k_2}=\dfrac{n_k(k_2)}{n}$. (\ref{linear equation}) is a linear equation of $A_k$ that can be solved analytically. If $|d_k|>2$, there are more than one ways to divide $d_k$ into two groups, each of them corresponds to  a merged MLE. Despite this, all of them are identical in terms of the statistical properties that will be discussed in Section \ref{mergedMLE}.
%
%Using and combining the strategies presented here, we can have various explicit estimators. For instance, one of the estimators used in \cite{ADV07} is one of them that divides $d_k$ into two groups and only considers the pairwise correlations between the members of the two groups. Therefore, although the estimator is proved to be a MLE in \cite{ Zhu11a} , it is not  exactly the same as the original MLE and we will discuss it further in Section \ref{mergedMLE}.

\section{Properties of Estimators} \label{section5}

 It is known that if a MLE is a function of the sufficient statistic, it is asymptotically unbiased, consistent and asymptotically efficient. Thus, the original MLE and
all of the estimators proposed in this paper have that properties. Apart from them,  whether there are other properties, such as unbiasedness, uniqueness, variance, and efficiency, that can be used to compare and evaluate the estimators are the main focus of this section. 

\subsection{Unbiasedness and Uniqueness of $Al(x)$ and $A_k(i)$}
This subsection is focused on the unbiasedness of the estimators in IBE and BWE although the statistic used by the latter is not minimal sufficient. For $Al_k(x), x \in \Sigma_k'$, we have the following theorem.

 \begin{theorem} \label{local maximum}
$Al_k(x)$ is a unbiased estimator.
\end{theorem}

\begin{IEEEproof}
Let $z_j, v_j \in d_k$ be the pass rate of $T(j)$ and let  $\overline{A_k}=\frac{N_k}{n}$ be the sample mean of $A_k$. Note that $z_j$ and $z_l, j, l \in d_k$ are independent from each other if $ j \neq l$. In addition, $z_j, v_j \in d_k$ is independent from $A_k$. Because of this, $x_k^i\prod_{v_j \in x} z_j$ is used to replace $\bigwedge_{v_j \in x} y_j^i$ in the following derivation since the latter is equal to $\prod_{v_j\in x} y_j^i$ that is equal to $x_k^i\prod_{v_j \in x} z_j$. We then have

\begin{eqnarray}
E(\widehat Al_k(x))&=&E\Big( \big (\frac{\prod_{v_j\in x} \hat\gamma_j}{\frac{I_k(x)}{n}}\big)^{\frac{1}{|x|-1}}\Big) \nonumber \\
&=& E\Big(\big(\frac{\prod_{v_j\in x} \frac{n_j(d_j)}{n}}{\frac{\sum_{i=1}^n \bigwedge_{v_j \in x} y^i_j}{n}}\big)^{\frac{1}{|x|-1}} \Big )\nonumber \\
&=& E\Big(\big(\frac{(\dfrac{N_k}{n})^{|x|}\prod_{v_j\in x} \frac{n_j(d_j)}{N_k}}{\frac{N_k}{n} \frac{\sum_{i=1}^{N_k}\prod_{v_j \in x} z_j}{N_k}}\big)^{\frac{1}{|x|-1}} \Big ) \nonumber \\
&=& E\Big(\frac{N_k}{n}\Big) E\Big(\big(\frac{\prod_{v_j\in x} \frac{1}{N_k}\sum_{i=1}^{N_k}{z_j}}{\sum_{i=1}^{N_k} \frac{1}{N_k}\prod_{v_j \in x} {z_j}}\big)^{\frac{1}{|x|-1}}\Big ) \nonumber \\
&=& E\Big (\overline{A_k}\Big). \nonumber
%\label{weighted mean}
\end{eqnarray}
%&=& E\Big((\dfrac{N_k}{n})^{|x|-1}\Big)E\Big(\dfrac{\prod_{v_j\in x} \overline{z_j}}{\prod_{v_j\in x} z_j}\Big)  \nonumber\\
%&=&E\Big ( \dfrac{N_k}{n}\Big) E\Big (\big( \prod_{v_j \in x} \dfrac{\overline{z_j}}{z_j} \big)^{\frac{1}{|x|-1}}\Big )   \nonumber \\
%&=&E\Big (\overline{A_k}^{|x|-1}\Big)\prod_{v_j \in x}(E(\dfrac{\overline{z_j}}{z_j}))  \nonumber\\

The theorem follows.
\end{IEEEproof}
%Note that there are only $N_k$ probes reaching node $k$. Then, $\frac{\sum_{i=1}^n \bigwedge_{v_j \in x} y^i_j}{n}$ is replaced by $\frac{N_k}{n} \frac{\sum_{i=1}^{N_k}\prod_{v_j \in x} z_j}{N_k}$ in the above derivation.
Given theorem \ref{local maximum}, we have the follow corollary.
 \begin{corollary} \label{global expect}
 $A_k(i)$ is a unbiased estimator.
 \end{corollary}
    \begin{IEEEproof}
According to theorem \ref{local maximum}, we have
\begin{eqnarray}
E(\widehat A_k(i))&=&E\Big(\overline{A_k}\Big)E\Big(\big(\frac{\sum_{x \in S(i)}\prod_{v_j\in x} \frac{1}{N_k}\sum_{i=1}^{N_k}{z_j}}{\sum_{x \in S(i)}\sum_{i=1}^{N_k} \frac{1}{N_k}\prod_{v_j \in x} {z_j}}\big)\big)^{\frac{1}{i-1}} \Big)\nonumber \\
&=&E\Big (\overline{A_k}\Big) \nonumber
\end{eqnarray}
%where $\overline{z_j}=\frac{n_j(d_j)}{N_k}$ is the sample mean of $z_j$.
\end{IEEEproof}

%Given theorem \ref{local maximum} and corollary \ref{global expect}, we have the following theorem to confirm the unbiasedness of the estimates obtained by either  $Al_k(x)$ or $A_k(i)$.
%
%
%  \begin{theorem} \label{local unbiased}
%The estimate obtained by either $Al_k(x)^{\frac{1}{|x|-1}}, x \in \Sigma_k'$ or by $A_k(i)^{\frac{1}{i-1}}, i \in \{2, \cdot\cdot,|d_k\}$ is an unbiased estimate of $A_k$.
%  \end{theorem}
%  \begin{IEEEproof}
%Bring the root operation into (\ref{weighted mean}) and  (\ref{global estimate}) , we have
%\begin{eqnarray}
%&&E(\widehat Al_k(x)^{\frac{1}{|x|-1}})=E\Big (\overline{A_k}\Big)=A_k, \mbox{and} \\
%&&E(\widehat A_k(i)^{\frac{1}{i-1}})=E\Big (\overline{A_k}\Big)=A_k.
%\label{local unbiased}
%\end{eqnarray}
%
% \end{IEEEproof}
Given $Al_k(x), x \in \Sigma_k'$ and $A_k(i)$ are unbiased estimators,
we can prove the uniqueness of $A_k(i)$.

\begin{theorem}
If
\[
\sum_{\substack{ x \in S_k(i)}} \prod_{v_j \in x} \hat\gamma_j < \sum_{x \in S_k(i)}\dfrac{I_k(x)}{n},
\]
there is only one solution in $(0,1)$ for $\widehat A_k(i), 2 \leq i \leq |d_k|$.
\end{theorem}
\begin{IEEEproof}
Since the support of $ A_k$ is in (0,1), we can reach this conclusion from (\ref{approximateestimator}).
\end{IEEEproof}

\subsection{Efficiency of  $Al_k(x)$, $Am_k(x)$, and the original MLE}
%Let $\prod_{v_j \in x} \beta_j = \psi_k(x)$. The likelihood function of $I_k(x)$ is
%\begin{eqnarray}
%L(A_k|I_k(x))=(A_k\psi_k(x))^{I_k(x)}(1- A_k\psi_k(x))^{n-I_k(x)}.
%\label{Al likelihood}
%\end{eqnarray}
Apart from  asymptotically efficiency stated previously for the MLEs using sufficient statistics, we are interested in the efficiency of the estimators proposed in this paper. Given (\ref{Al likelihood1}), we have the following theorem for the Fisher information of an observation, $y$, on the estimators in IBE, i.e. $Al_k(x), x \in \Sigma_k'$.
\begin{theorem} \label{Al fisher}
The Fisher information of $y$ on $Al_k(x), x \subset d_k$ is equal to $ \dfrac{\psi_k(x)}{A_k (1-A_k \psi_k(x))}$.
\end{theorem}
\begin{IEEEproof}
Considering $I_k(x)=y$ is the observation of the receivers attached to $x$, we have the following as the likelihood function of the observation:
\begin{equation}
{\cal L}(A_k, Y_k(x), y)=y\log (A_k\psi_k(x))+(1-y)\log(1-A_k\psi_k(x)).
\label{Al likelihood for single}
\end{equation}
Differentiating (\ref{Al likelihood for single}) {\it wrt} $A_k$, we have
\begin{eqnarray}
\dfrac{\partial {\cal L}(A_k, Y_k(x), y)}{\partial A_k}=\dfrac{y}{A_k}-\dfrac{(1-y)\psi_k(x)}{1-A_k\psi_k(x)} \nonumber
\end{eqnarray}
We then have
\begin{eqnarray}
\dfrac{\partial^2 {\cal L}(A_k, Y_k(x), y)}{\partial A_k^2}&=& -\dfrac{y}{A_k^2}-\dfrac{(1-y)\psi_k(x)^2}{(1-A_k\psi_k(x))^2} \nonumber
\end{eqnarray}
If ${\cal I}(Al_k(x)|y)$ is used to denote the Fisher information of observation $y$ for $A_k$ in $Al_k(x)$, we  have
\begin{eqnarray}\label{fisher}
{\cal I}(Al_k(x)| y)&=&-E(\dfrac{\partial^2 {\cal L}(A_k, Y_k(x), y)}{\partial A_k^2}) \nonumber \\
&=&\dfrac{E(y)}{A_k^2}+\dfrac{E(1-y)\psi_k(x)^2}{(1-A_k\psi_k(x))^2} \nonumber \\
&=&\dfrac{\psi_k(x)}{A_k (1-A_k \psi_k(x))}
%&=& \dfrac{1}{i-1}A_k \prod_{v_j \in x} \beta_j
\end{eqnarray}
that is the information provided by $y$ for $A_k$.
\end{IEEEproof}
Given (\ref{fisher}), we have a formula to compute the Fisher information of the original MLE and the estimators in RSE. In order to use a formula for all of them, let $\beta_k(d_k)=\beta_k$. Then, we have the following corollary.
\begin{corollary} \label{MLE fisher}
The Fisher information of observation $y$ for $A_k$ in the original MLE and $Am_k(x), x \subseteq d_k$ is equal to
\begin{equation}
 \dfrac{\beta_k(x)}{A_k (1-A_k \beta_k(x))}, \mbox{     } x \subseteq d_k.
 \label{MLE fisher equ}
 \end{equation}
\end{corollary}
\begin{IEEEproof}
Replacing $n_k(d_k)$ or $n_k(x)$ by $y$ and replacing $n-n_k(d_k)$ or $n-n_k(x)$ by $1-y$ from (\ref{likelihood}) and (\ref{likelihood RSE}), respectively, and then using the same procedure as that used in the proof of theorem \ref{Al fisher}, the corollary follows.
\end{IEEEproof}
Because of the similarity between (\ref{fisher}) and (\ref{MLE fisher equ}), the two equations have the same features in terms of support, singularity, and maximum. After
eliminating the singular points, the support of $A_k$ is in $(0,1)$ and the support of $\beta_k(x)$ (or $\psi_k(x)$) is in $[0, 1]$.
Both (\ref{fisher}) and (\ref{MLE fisher equ}) are convex functions in the support and reach the maximum at the points of  $A_k \rightarrow 1, \beta_k(x) =1$ (or ($\psi_k(x)=1$) and $A_k\rightarrow 0, \beta_k(x)=1$ (or ($\psi_k(x)=1$).
%The formulae show that the efficiency of $Al_k(x)$ or the MLE is determined by the ratio of $\psi_k(x)$ to $A_k(1-A_k\psi_k(x)$.
Given $A_k$, (\ref{MLE fisher equ}) is  a monotonic increase function of $\beta_k(x)$ whereas (\ref{fisher}) is a monotonic increase function of $\psi_k(x)$. 

Despite the similarity between (\ref{fisher}) and (\ref{MLE fisher equ}),  $Am_k(x)$ and $Al_k(x)$ react differently if $x$ is replaced by $y, x \subset y$ in terms of efficiency that leads to two corollaries, one for each of them.
\begin{corollary}
$Am_k(y)$ is more efficient than $Am_k(x)$ if $x \subset y$.
\end{corollary}
\begin{IEEEproof}
If $x \subset y$, $\beta_k(x) \leq \beta_k(y)$ and then we have the corollary.
\end{IEEEproof}
For $Al_k(x)$, we have
\begin{corollary} 
The efficiency of $Al_k(x), x \in \Sigma_k'$ forms a partial order that is identical to that formed on the inclusion of the members in $\Sigma_k'$, where the most efficient estimator must be one of the $Al_k(x), x \in S_k(2)$ and the least efficient one must be $Al_k(d_k)$.
\label{Al corollary}
\end{corollary}
\begin{IEEEproof}
According to Theorem \ref{Al fisher}, the efficiency of $Al_k(x)$ is determined by $\psi_k(x)$, where $\psi_k(x)=\prod_{v_j \in x} \alpha_j \beta_j$. If $x \subset y$, we have
 \begin{eqnarray}
 \psi_k(y)&=&\prod_{v_j \in y} \alpha_j \beta_j \nonumber \\
 &=&\psi_k(x)\prod_{v_j \in (y \setminus x)} \alpha_j \beta_j \nonumber \\
 &<& \psi_k(x). \nonumber
 \end{eqnarray}
 Therefore, the order of the efficiency of $\{Al_k(x): x \in \Sigma_k'\}$ is identical to the order of the inclusion in $\Sigma_k'$, where $\{x: x \in S_k(2)\}$ are the members of $\Sigma_k'$ that have the minimal number of elements.
 In contrast to $\{\psi_k(x): x \in S_k(2)\}$, $\psi_k(d_k)\leq \psi_k(x)$ since $\forall x, x \in \Sigma_k' \to x \subseteq d_k$. Then, the corollary follows.
\end{IEEEproof}

\subsection{Variance of $Al_k(x)$, $Am_k(x)$, and the original MLE}
The estimator specified by (\ref{minc}),  $Am_k(x)$, and $Al_k(x)$ are of MLEs that have different focuses on the observations obtained by receivers. Despite the difference between them, they share a number of features, including likelihood function and efficient equation. In addition, the variances of them are expressed by a function showing the connection between $A_k$ and the pass rate of the subtree(s) connecting node $k$ to the receivers. Let $mle$ denote all of them and then we have a theorem for the variances of the estimators in $mle$.

\begin{theorem} \label{Al variance}
The variances of the estimators in $mle$ equal to
\begin{equation}
var(mle)=\dfrac{A_k (1-A_k\delta_k(x) )}{\delta_k(x)}, \mbox{  } x \subseteq d_k
\label{Al variance1}
\end{equation}
where $\delta_k(x)$
\begin{eqnarray}
\delta_k(x)=\begin{cases} \beta_k(x), & \mbox{for the original MLE and }  Am_k(x); \\  \psi_k(x), & \mbox{for } Al_k(x). \end{cases} \nonumber
\end{eqnarray}
\end{theorem}
\begin{IEEEproof}
The passing process described by (\ref{Al likelihood1}) is a Bernoulli process that falls into the exponential family and satisfies the regularity conditions presented in \cite{Joshi76}. Thus, the variance of  an estimator in $mle$ reaches the Cram\'{e}r-Rao bound that is the reciprocal of the Fisher information.
\end{IEEEproof}
(\ref{Al variance1}) can be written as
\begin{eqnarray}
\frac{A_k}{\delta_k(x)}-A_k^2 \nonumber
\end{eqnarray}
which shows:
\begin{enumerate}
 \item the estimates obtained by an estimator spread out more widely than that obtained by direct measurement. The wideness is determined by $\delta_k(x)$, the pass rate of the subtrees connecting node $k$ to the observers. If $\delta_k(x)=1$, there is no further spread-out than that obtained by direct measurement. Otherwise, the variance increases as the decreases of $\delta_k$ and in a super linear fashion.
\item
the variance of the estimates obtained by an estimator is monotonically increasing as the depth of the subtree rooted at node $k$ since the pass rate of a subtree decreases as its depth, i.e., the pass rate of an $i$-level tree is larger than that of the $i+1$-level one that is extended from the $i$-level one;
%\item the variance of the estimates  obtained by an estimator in $mle$ is a monotonically decreasing function of $\delta_k(x)$. 
\end{enumerate}
The two points agree with some of the experiment results reported previously, such as the dependency of variance on topology reported in \cite{CDHT99}. Note than the variance of $Am_k(x)$ can be the same as that of $Am_k(y), x \subset y$ if $\beta_k(x)=\beta_k(y)$. So does $Al_k(x)$. In other words, if the probes observed by $R(j), v_j \in (y\setminus x)$ are included in that observed by $R(i), i \in x$, the estimate obtained by $Am_k(x)$ is the same as that obtained by $Am_k(y)$.
%On the other hand, given $T$, if we move from node $k$ downward along a path that has the lowest pass rate to a descendant, say $j$, $A_j$ is equal or smaller than $A_k$ and $\beta_j$ may be bigger than $\beta_k$. If so, the variance of the estimates obtained for $A_j$ can be different from that for $A_k$.

\subsection{Efficiency and Variance of BWE}

As stated, the estimate obtained by $A_k(i)$ is a type of the arithmetic mean of $Al_k(x), x \in S_k(i)$ that has the same advantages and disadvantages as the arithmetic mean. Thus, $A_k(i)$ is more robust and efficient than that of $Al_k(x), x \in S(i)$ since the former considers more probes than the latter in estimation although some of the probes may be considered more than once. Because of this, (\ref{fisher}) cannot be used to evaluate the efficiency of an estimator in BWE. Despite this, we can put a range for the information obtained by $A_k(i)$ that is
 \begin{eqnarray}
\frac{\psi_k(x)}{A_k(1-A_k\psi_k(x))} \leq {\cal I}(A_k(i)|y) \leq  {d_k \choose i} \frac{\psi_k(x)}{A_k(1-A_k\psi_k(x))}.  \mbox{    } |x|=i. \nonumber
 \end{eqnarray}
In addition, $A(i)$ is at least as efficient as $A(i+1)$ and the variance of $A(i)$ is at least as small as that of $A(i+1)$ since $\sum_{x \in S_k(i)} I_k(x) \leq \sum_{x \in S_k(i+1)} I_k(x)$.
 %Although Godambe information \cite{Godambe91} can be used here, we choose an intuitive approach to rank the efficiency of the estimators in BWE for simplicity.
% Nevertheless, as a special arithmetic mean of $Al_k(x), |x|=i$, $A_k(i)$ shares many features of $Al_k(x)$.
%The results presented in this paper show that the efficiency, variance, etc, of an estimator are determined by the number of probes observed and considered by the estimator without redundancy. This principle is also applicable to $A_k(i)$ and then we can conclude that
%One of the features is that the efficiency is determined by the number of probes actually used in estimation.
\subsection{Variance of the merged MLE}
\label{mergedMLE}

Although the results reported in \cite{ADV07} show the merged MLE
 performs almost identical to the original MLE in simulations and it was proved to be a MLE in \cite{Zhu11a}, the estimator has been overlooked since the authors doubt the estimator may have a higher variance. However, there is no statistical evidence to support the doubt. In addition, there are a number of issues about the merged MLE that have not been addressed, such as how to divide $d_k$ into groups in order to achieve the best performance and whether it performs the same regardless of  how to divide the multicast subtrees rooted at node $k$ into exclusive groups.  Given theorem \ref{Al variance}, we are able to answer the question and have the following corollary.

\begin{corollary} \label{merged MLE}
The variance of the estimates obtained by the merged MLE, regardless of the strategy used to divide $d_k$ into groups, is equal to
\begin{equation}
var(merged \mbox{ }mle)=\dfrac{A_k (1-A_k\beta_k)}{\beta_k}. \nonumber
\end{equation}
\end{corollary}
\begin{IEEEproof}
If  $d_k$ is divided into a number of exclusive groups, where $\zeta_k$ is used to denote the groups, a multicast tree is structured that has a virtual link to connect $v_0$ to $v_k$ and has $|\zeta_k|$ virtual links to connect $v_k$ to $|\zeta_k|$ virtual receivers. The statistics of the virtual receivers are obtained by (\ref{nk2}).  Then, the variance of the estimates obtained by the merged MLE  is equal to
 \begin{equation}
var(merged\mbox{ }mle)=\dfrac{A_k (1-A_k\beta_k(\zeta_k))}{\beta_k(\zeta_k)}, \nonumber
\end{equation}
where $\beta_k(\zeta_k)$  denotes the pass rate of the newly structured subtree rooted at node $k$, according to theorem \ref{Al variance}. To prove $\beta_k(\zeta_k)=\beta_k$, it is equal to prove
\begin{equation}
1-\beta_k=1-\beta_k(\zeta_k),
\label{betakequal}
\end{equation}
where the RHS of (\ref{betakequal}) can be written as
\begin{equation}
1-\beta_k(\zeta_k)=\prod_{q \in \zeta_k} (1-\beta_k(q))
\label{subgroup0}
\end{equation}
since
\begin{equation}
1-\beta_k(q)=\prod_{v_j \in q}(1-\frac{\gamma_j}{A_k}). 
\label{subgroup1}
\end{equation}
Using the RHS of (\ref{subgroup1}) to replace the terms on the RHS of (\ref{subgroup0}), we have
\begin{equation}
1-\beta_k(\zeta_k)=\prod_{v_j \in d_k}(1-\frac{\gamma_j}{A_k}). \nonumber
\end{equation}
We then have
\begin{equation}
\beta_k=\beta_k(\zeta_k) \nonumber
\end{equation}
and the corollary follows.
\end{IEEEproof}
\subsection{Example}
We use an example to conclude this section that illustrate the differences of the variances obtained from the estimates of four estimators. The four estimators are: direct measurement, the original MLE, $Al_k(x), |x|=2$ and $Al_k(d_k)$, respectively. The setting used here is identical to that presented in \cite{DHPT06}, where node $k$ has three children with a pass
rate of $\alpha, 0 < \alpha \leq 1$, and the pass rate from the root to node $k$ is also equal to $\alpha$.  Using (\ref{Al variance1}), we have the variances of them that are presented below:
\begin{enumerate}
\item $\alpha-\alpha^2$,
\item $\frac{1}{3(1-\alpha)+\alpha^2}-\alpha^2$,
\item $\frac{1}{\alpha}-\alpha^2$, and
\item $\frac{1}{\alpha^2}-\alpha^2$.
\end{enumerate}
The difference between them becomes obvious as $\alpha$ decreases from $1$ to  $0.99$, where the variances of the four estimators change from 0 to 0.01, 0.01, 0.03, and 0.04, respectively. The variance of $Al_k(d_k)$ is 4 times of that of the original MLE that is significantly different from that obtained in \cite{DHPT06}. Although the variances are decreased as the number of probes multicasted, the ratio between them remains.

\section{Summary and Conclusion}\label{section7}

Although loss tomography has been studied for quite some time and a number of estimators have been proposed, there has been a lack of statistical analyses to the finite sample properties of an estimator, including that proposed in \cite{CDHT99}. Without the properties, simulations have been widely used to evaluate the performance of an estimator. However,  it is difficult to interpret the simulation results and find the weakness of an estimator since there are too many random factors that can influence the results.
This paper aims to fill the gaps that starts from investigating the the fundamental principle  used by the MLE proposed in \cite{CDHT99} and ends with a number of theorems and corollaries that cover the most important statistical properties of the MLEs, including that proposed in \cite{CDHT99}. The investigation unveils that all of the estimators, including the MLEs, rely on the correlations between predictors and observations to estimate the loss rate of a link from end-to-end measurement. The investigation also finds a number of weaknesses  within the previous works, such as the lack of a model to connect observation to the parameter to be estimated. In this regard, a two-segment model, where the first segment is from the source to the common ancestor of a group receivers and the second segment is from the ancestor to the receivers, is  proposed to model the probing process.

Using the model, we are able to identify a number of the minimal sufficient  statistics for various likelihood functions that subsequently lead to a number of explicit estimators. To evaluate the estimators, the statistical properties of the proposed estimators, in particular the unbiasedness, efficiency and variance, are presented that ensures the mean of the estimates obtained from them is equal to the parameter to be estimated and the variances of the estimates decrease as the pass rate of the second segment. In general, the variance is inversely proportional to the pass rate of the second segment. If the pass rate of the second segment is equal to 1, the variance of the estimates is equal to the variance of the first segment. Thus, the pass rate of the second segment determines the quality of an estimator. If the pass rate of an estimator is higher than that of another,  the variance of the former is smaller than that of the latter. This shows the dependency between the segments on the path disseminating probes from source to receivers, which is the nature of inference. The properties are also reflected on the simulation result conducted by other researchers, such as those presented in \cite{CDHT99, LHATSBY15}. 

Apart from the above, the model plays an important role to correct a cognition that has been widely used in the literature  \cite{CDHT99, DHPT06, ADV07, Zhu11a, LHATSBY15}, which claims the degree of the polynomial used for the most efficient MLE is one less than the number of descendants attached to the path of interest. In this paper, the merged MLEs are  proved to be one of the most efficient MLEs and can be expressed by a linear function of $A_k$. That shows  the second segment in the model acts exactly the same as the first segment that delivers probes from one end to another  regardless of the number of paths and the number of receivers. If each  of the segments is modelled as a Bernoulli process and we are interested in the loss rate of the first segment, there is no need to distinguish the subtrees within the second segment in the likelihood equation, in particular if multicast is used to disseminate probes.

The statistical results presented in this  paper can be applied to other areas of network tomography, especially if a Bernoulli model is used to describe the behaviour of the characteristics of interest. For instance,  the model used and the theorems obtained in this paper can be extended to estimate the loss rates of the links in a network with the general topology.

\bibliography{../globcom06/congestion}

\begin{thebibliography}{10}

\bibitem{YV96}
Y.~Vardi, ``Network tomography: Estimating source-destination traffic
  intensities from link data,'' {\em Journal of Amer. Stat. Association},
  vol.~91, no.~433, 1996.

\bibitem{CDHT99}
R.~C\'{a}ceres, N.~Duffield, J.~Horowitz, and D.~Towsley, ``Multicast-based
  inference of network-internal loss characteristics,'' {\em IEEE Trans. on
  Information Theory}, vol.~45, 1999.

\bibitem{CDMT99}
R.~C\'{a}ceres, N.~Duffield, S.~Moon, and D.~Towsley, ``{Inference of Internal
  Loss Rates in the MBone },'' in {\em IEEE/ISOC Global Internet'99}, 1999.

\bibitem{CDMT99a}
R.~C\'{a}ceres, N.~Duffield, S.~Moon, and D.~Towsley, ``Inferring link-level
  performance from end-to-end multicast measurements,'' tech. rep., University
  of Massachusetts, 1999.

\bibitem{CN00}
M.~Coates and R.~Nowak, ``Unicast network tomography using {EM} algorithms,''
  Tech. Rep. TR-0004, Rice University, September 2000.

\bibitem{XGN06}
B.~Xi, G.~Nichailidis, and V.~Nair, ``Estimating network loss rates using
  active tomography,'' {\em Journal of the American Statistical Association},
  vol.~101, no.~476, 2006.

\bibitem{BDPT02}
T.~Bu, N.~Duffield, F.~Presti, and D.~Towsley, ``Network tomography on {General
  Topologies},'' in {\em SIGCOMM 2002}, 2002.

\bibitem{ADV07}
V.~Arya, N.~Duffield, and D.~Veitch, ``Multicast inference of temporal loss
  characteristics,'' {\em Performance Evaluation}, vol.~9-12, 2007.

\bibitem{DHPT06}
N.~Duffield, J.~Horowitz, F.~L. Presti, and D.~Towsley, ``Explicit loss
  inference in multicast tomography,'' {\em IEEE trans. on Information Theory},
  vol.~52, no.~8, 2006.

\bibitem{ZG05}
W.~Zhu and Z.~Geng, ``Bottom up inference of loss rate,'' {\em Journal of
  Computer Communications}, vol.~28, no.~4, 2005.

\bibitem{GW03}
D.~Guo and X.~Wang, ``Bayesian inference of network loss and delay
  characteristics with applications to tcp performance predication,'' {\em IEEE
  trans. on Signal Processing}, vol.~51, no.~8, 2003.

\bibitem{LY03}
G.~Liang and B.~Yu, ``Maximum pseudo likelihood estimation in network
  tomography,'' {\em IEEE trans. on Signal Processing}, vol.~51, no.~8, 2003.

\bibitem{TCN03}
Y.~Tsang, M.~Coates, and R.~D. Nowak, ``Network delay tomography,'' {\em IEEE
  trans. on Signal Processing}, vol.~51, no.~8, 2003.

\bibitem{PDHT02}
F.~L. Presti, N.~Duffield, J.~Horowitz, and D.~Towsley, ``Multicast-based
  inference of network-internal delay distributions,'' {\em IEEE/ACM trans. on
  Networking}, vol.~10, 2002.

\bibitem{SH03}
M.-F. Shih and A.~Hero, ``Unicast-based inference of network link delay
  distribution with finite mixture models,'' {\em IEEE trans. on Signal
  Processing}, vol.~51, no.~8, 2003.

\bibitem{LGN06}
E.~Lawrence, G.~Michailidis, and V.~N. Nair, ``Network delay tomography using
  flexicast experiments,'' {\em Journal of Royal Statistical Society, Series
  B}, vol.~68, no.~5, 2006.

\bibitem{LHATSBY15}
C.~Liu, T.~He, A.~Swami, D.~Towsley, T.~Salonidis, A.~I. Bejan, and P.~Yu,
  ``Multicast vs. unicast for loss tomography on tree topologies,'' in {\em
  2015 IEEE Military Communications Conference}, 2015.

\bibitem{HLATSBY15}
T.~He, C.~Liu, A.~Swami, D.~Towsley, T.~Salonidis, A.~I. Bejan, and P.~Yu,
  ``Fisher information-based experiment design for network tomography,'' in
  {\em Proceedings of the 2015 ACM SIGMETRICS International Conference on
  Measurement and Modeling of Computer Systems}, pp.~389--402, 2015.

\bibitem{HBB00}
K.~Harfoush, A.~Bestavros, and J.~Byers, ``Robust identification of shared
  losses using end-to-end unicast probes,'' in {\em Technical Report
  BUCS-2000-013}, Boston University, 2000.

\bibitem{Lindsay88}
B.~C. Lindsay, ``Composite likelihood method,'' {\em Contemporary Mathematics},
  vol.~80, 1988.

\bibitem{Zhu11a}
W.~Zhu, ``An efficient loss rate estimator in multicast tomography and its
  validity,'' in {\em IEEE International Conference on Communication and
  Software}, 2011.

\bibitem{RM96}
R.~Mittelhammer, {\em Mathematical Statistics for Economics and Business},
  vol.~78.
\newblock Springer, 1996.

\bibitem{AS10}
R.~B. J.~T. Allenby and A.~Slomson, {\em How to Count: An Introduction to
  Combinatorics, Discrete Mathematics and Its Applications}.
\newblock CRC Press, second~ed., 2010.

\bibitem{Besay74}
J.~Besag, ``Spatial interaction and the statistical analysis of lattice system
  (with discussion),'' {\em Journal of Royal Statistical Society}, vol.~36,
  1974.

\bibitem{Joshi76}
V.~M. Joshi, ``On the attainment of the cramer-rao lower bound,'' {\em Ann.
  Statist.}, vol.~4, no.~5, pp.~998--1002, 1976.

\end{thebibliography}

%\bibliography{./congestion}
%&=& E\Big(\dfrac{\prod_{v_j\in x} \dfrac{n_j(d_j)}{n}}{\dfrac{\sum_{i=1}^n \bigwedge_{v_j \in x} y_j^i}{n}}\Big ) \nonumber \\
%&=& E\Big(\dfrac{\prod_{v_j\in x}\Big(\dfrac{\dfrac{n_i(1)}{N_k}}{\dfrac{n}{N_k}}\Big)}{\dfrac{\sum_{i=1}^n f(z_x)^i \prod_{v_j \in x} z %_j^i}{n}} \Big ) \nonumber \\
%&=&  E\Big(\dfrac{\prod_{v_j \in x} \dfrac{n_j(d_j)}{N_k}}{\prod_{v_j \in x} \dfrac{n}{N_k}\cdot \dfrac{1}{n}\cdot n \cdot A_k \prod_{v_j \in %x}  \dfrac{n_j(d_j)}{N_k}} \Big ) \nonumber \\

\end{document}